\documentclass[]{article}
\usepackage{times}
\usepackage{graphicx}
\usepackage{amsmath,amssymb,bm}
\usepackage{epsf}

\newcommand{\vect}[1]{\bm{#1}} 

\begin{document}

\title{Inertial and retardation effects for dislocation interactions}

\author{L. Pillon and C. Denoual \\ CEA, DAM, DIF, Arpajon, F-91297, France}

\date{\small \emph{To cite this Article}: Pillon, L. and Denoual, C. (2009) `Inertial and retardation effects for dislocation interactions', Philosophical Magazine, 89:2, 127--141}

\maketitle

\section*{Abstract}
A new formulation for the equation of motion of interacting dislocations is derived. From this solution it is shown that additional coupling forces, of kinetic and inertial origin, should be considered in Dislocation Dynamics (DD) simulations at high strain rates. A heuristic modification of this general equation of motion enables one to introduce retardation into inertial and elastic forces, in accordance with a progressive rearrangement of fields through wave propagation. The influence of the corresponding coupling terms and retardation effects are then illustrated in the case of dislocation dipolar interaction and coplanar annihilation. Finally, comparison is made between the modified equation of motion and a precise numerical solution based on the Peierls-Nabarro Galerkin method. Good agreement is found between the Peierls-Nabarro Galerkin method and the EoM including retardation effects for a dipolar interaction. For coplanar annihilation, it is demonstrated that an unexpected mechanism, involving a complex interplay between the core of the dislocations and kinetics energies, allows a renucleation from the completely annihilated dislocations. A description of this phenomenon that could break the most favourable reaction between dislocations is proposed.

\emph{keywords}:
dislocation interactions, elastic waves, plasticity of metals, inertial effect

\section{Introduction}
During the last decade, the understanding of crystal plasticity has been considerably improved with the rapid development of Dislocation Dynamics (DD) simulations. These simulations have indeed the capacity to quantitatively predict the behaviour of micrometric samples from the modelling of the motion of discrete dislocations and interactions at the elementary scale \cite{Devincre_Kubin_97,Madec_etal_03,Bulatov_etal_06,Madec_etal_08,Devincre_etal_08}. Up to now, these simulations have been mainly devoted to low--strain rate deformations. In this case, dislocations motion is well described as a steady--state motion, the inertial forces being negligible. Recent extensions toward more dynamic loadings (e.g. shock loadings) pointed out that inertial effects can be important, notably to overcome obstacles like other dislocations or defects \cite{Bitzek_Gumbsch_04,Weygand_04,Bitzek_Weygand_Gumbsch_04,Bitzek_Gumbsch_05,Wang_etal_07}. This is why, to extend the capacity of DD simulations to high--strain rates, works have been dedicated to the complex problem of dynamic equation of motion \cite{Eshelby_53,Clifton_Markenscoff_81,Pillon_etal_07}.

Inertial effects for a single dislocation stem from the modification of the amount of energy, both elastic and kinetic, that follows change in the dislocation velocity. To balance such variations, supplementary work has to be done by the so-called inertial force. A simple estimation of this force relies on the hypothesis of steady-state stress and velocity fields around a dislocation \cite{Hirth_Zbib_Lothe_98}. However, rearrangement of the fields through wave emission has been shown to be critical in order to quantitatively describe the effect of inertia \cite{Eshelby_53,Pillon_etal_07}.

Wave propagation naturally leads to a retardation of the interaction between dislocations that may have a very strong influence on shock loadings. In conventional DD simulations, the change of elastic forces due to dislocation motion is considered instantaneous, without being limited by the speed of the sound waves. Therefore, the dislocations moving behind a shock front artificially alter the stress in front of it. To avoid such unphysical propagation, retardation effects have to be considered in the dislocation-dislocation interaction forces.

In this paper, we propose a modelling for inertia and retardation effects in the framework of the equations of motion developed for the Dislocation Dynamic method. Two Equations of Motion (EoM) for interacting dislocations are proposed. In a first section, the solution of instantaneously updated fields allows the definition of all the terms appearing in the EoM. In this first equation, the overall kinetic energy does not reduce to the sum of the kinetic energies of each isolated dislocation. This induces additional coupling terms between dislocations of kinetic and inertial nature, the importance of which will be discussed. The second section is devoted to a heuristic modification of the first EoM in which retardation effects are included in the inertial and elastic interaction forces. In the third section, the results from both EoMs are compared to the results of a full--dynamic and more fundamental method, the Peierls-Nabarro Galerkin method. This method corresponds exactly to the theoretical framework used for the proposed EoM, excepted for the treatment of acoustic waves, now exactly resolved. This comparison is concluded by a discussion on the influence of inertia, on retarded effects and on the coupling terms accounted for in a dynamic EoM, for a dipolar interaction and a coplanar annihilation. In the latter configuration, an unusual mechanism of renucleation from the annihilated dislocations is depicted. A discussion of this phenomenon is proposed in the last section and emphasizes the complex interaction between kinetic and core energies of the dislocations.

\section{Instant Equation of Motion}
\label{sec:EoM_no_retardation}
Many mechanisms involved during dislocations interactions can be investigated with the simplistic problem of two attractive parallel dislocations of opposite signs (noted $\alpha$ and $\beta$). In this model, fields around dislocations are supposed to be modified everywhere in a time interval very short compared to the time needed by the acoustic waves to propagate in the solid. Therefore, at each time, the fields are close to stationary solutions and are only function of dislocation location (e.g. $x^\alpha$) and velocity (e.g. $v^\alpha$) \cite{Weertman_61}. The total energy of the system $E$ is obtained from the overall velocity fields $\dot{\vect{u}} = \dot{\vect{u}}^{\alpha} + \dot{\vect{u}}^{\beta}$ and stress fields $\vect{\sigma} = \dot{\vect{\sigma}}^{\alpha} + \dot{\vect{\sigma}}^{\beta}$, obtained by summing up the contributions of the two dislocations : 
\begin{equation}
E = e^{\alpha\alpha} + e^{\beta\beta} + 2e^{\alpha\beta}+ k^{\alpha\alpha} + k^{\beta\beta} + 2k^{\alpha\beta}
\end{equation}
where $e$ denotes an elastic energy and $k$ stands for a kinetic one:
\begin{eqnarray}
\label{e:def_e_k_1}
e^{\alpha\!\beta} 
&=&
\frac{1}{2} \int_\Omega  
      \vect{\sigma}^{\alpha}(\vect{r}-\vect{x}^{\alpha},v^{\alpha}) : 
      \vect{C}^{-1} : 
      \vect{\sigma}^{\beta}(\vect{r}-\vect{x}^{\beta},v^{\beta}) {\rm d}\vect{r}  \\
\label{e:def_e_k_2}
k^{\alpha\!\beta}  
&=& 
\frac{1}{2} \rho \int_\Omega
          \dot{\vect{u}}^{\alpha}(\vect{r}-\vect{x}^{\alpha},v^{\alpha}) 
          \cdot 
          \dot{\vect{u}}^{\beta} (\vect{r}-\vect{x}^{\beta},v^{\beta})   {\rm d}\vect{r}
\end{eqnarray}
where $\vect{C}$ is the stiffness tensor.  Terms noted by a double superscript $\square^{\alpha\alpha}$ or $\square^{\beta\beta}$ are related to isolated dislocations whereas mixed ones represent the cost of the interaction. With a hypothesis of instantaneous updated fields, the energy is only function of locations $x^\alpha$ and $x^\beta$ and velocities $v^\alpha$ and $v^\beta$. In the case of two dislocations of opposite signs and of symmetrical trajectories, energy conservation leads to the EoM (given in the following for the dislocation $\alpha$):
\begin{equation}
\label{e:EOM_Instant}
-2\frac{\partial (e^{\alpha\beta} + k^{\alpha\beta} )}{\partial x^{\alpha}}  =
\frac{\dot{v}^{\alpha}}{v^{\alpha}} \frac{\partial}{\partial v^{\alpha}}
\left[  
 e^{\alpha\alpha} + k^{\alpha\alpha} + 2e^{\alpha\beta} + 2k^{\alpha\beta} 
\right]
\end{equation}
Terms in the right-hand-side of equation~(\ref{e:EOM_Instant}) are proportional to acceleration and are of inertial nature whereas terms in the left-hand-side represent interaction forces. A brief description of each terms will be given now. We note $F^{\mathrm{E}}_\mathrm{i} = -2\partial e^{\alpha\beta}  / \partial x^{\alpha}$ the classical elastic interaction force and $F^{\mathrm{K}}_\mathrm{i} = -2\partial k^{\alpha\beta} / \partial x^{\alpha}$ a kinetic interaction force (the subscript `i' stands for ``instantaneous''). Inertia is made of two terms, the first one (termed self--inertial, SI) $F^{\mathrm{SI}}_\mathrm{i}$ characterizes the inertia of a single and isolated dislocation, as already defined by Hirth \emph{et al.} \cite{Hirth_Zbib_Lothe_98}:
\begin{equation}
\label{eq:edm_hirth} 
F^{\mathrm{SI}}_\mathrm{i} 
= \frac{\dot{v}^{\alpha}}{v^{\alpha}} \;\frac{\partial (e^{\alpha\alpha} + k^{\alpha\alpha} )}{\partial v^{\alpha}} 
= m\left[v^{\alpha}\right]\dot{v}^{\alpha}
\end{equation}
where $m(v)$ is termed the instant mass of a single dislocation. This mass is a complex function of velocity and becomes unbounded for $v$ approaching the shear wave velocity due to the divergence of strain and velocity fields (given in the following for an edge dislocation):
\begin{equation}
\label{eq:mass_hirth} 
m(v) = m_{s,0}\left(\frac{c_\text{S}}{v}\right)^4 
\left[ 
-8\gamma_{\rm L} 
-20\gamma_{\rm L}^{-1}
+4\gamma_{\rm L}^{-3} 
+7\gamma_{\rm S} 
+25\gamma_{\rm S}^{-1}
-11\gamma_{\rm S}^{-3}
+3\gamma_{\rm S}^{-5} 
\right] 
\end{equation}
with the mass of a screw dislocation at rest $m_{s,0}=\frac{\mu b^2}{4 \pi c_\text{S}^{2}}{\rm ln}\left[\frac{R}{r_{\rm 0}} \right]$, depending on the shear modulus $\mu$ and on the shear wave speed $c_\text{S}$ and $\gamma_{\rm L,S}=\left(1-v^2/c_{\rm L,S}^2\right)^{1/2}$, with the longitudinal wave speed $c_\text{L}$. The instant mass depends on the size $R$ of the domain in which the strain and velocity fields are supposed to be adapted to the present dislocation velocity. The parameter $r_{\rm 0}$ is a cut-off radius usually chosen to be equal to $b$.

The second inertial term $F^{\mathrm{II}}_\mathrm{i} = 2(\dot{v}^{\alpha} / v^{\alpha}) \;\partial (e^{\alpha\beta} + k^{\alpha\beta} )/\partial v^{\alpha}$, represents an ``inter--inertial'' (II) force. It can be noted that the equation of motion (\ref{e:EOM_Instant}) (termed in the following as the ``instant'' EoM), which now reads
\begin{equation}
\label{e:EOM_synthetic}
F^{\mathrm{E}}_\mathrm{i} + F^{\mathrm{K}}_\mathrm{i} = F^{\mathrm{SI}}_\mathrm{i} +  F^{\mathrm{II}}_\mathrm{i}
\end{equation}
contains two coupling terms ($F^{\mathrm{K}}_\mathrm{i}$ and $F^{\mathrm{II}}_\mathrm{i}$), usually not considered in studies of inertial effects for high velocity dislocations \cite{Roos_etal_01_a,Pillon_etal_06,Wang_etal_07}. However, the influence of these terms has not been shown to be negligible. In particular, when two dislocations superimpose (for example when a junction is created), the overall energy is not reduced to the sum of the individual energies. Coupling energies $e^{\alpha\beta}$ and $k^{\alpha\beta}$, from which the forces $F^{\mathrm{K}}_\mathrm{i}$ and $F^{\mathrm{II}}_\mathrm{i}$ are derived, can not been neglected in general and could \emph{a priori} play a role. 

An estimation of the kinetic and elastic interaction energies is obtained numerically for two straight and parallel dislocations of opposite signs and velocities. By noting that with the considered symmetries $\vect{x}_\alpha = -\vect{x}_\beta$, equations~(\ref{e:def_e_k_1}) and (\ref{e:def_e_k_2}) reduce to a simple convolution product performed with a fast Fourier transform. Relativistic stationary fields for stress, strain, velocity of a finite core dislocation are considered~\cite{Weertman_61}. 

The hypothesis of instantaneously updated fields is however a quite strong assumption since any changes in the velocity of the dislocation cannot be propagated more rapidly than the shear or longitudinal wave celerity. This is particularly true for high strain rate loadings in which a significant dislocation motion can occur during this propagation time. Hence, in the following section we construct a modified EoM in which retarded effects are now considered.

\section{Retarded Equation of Motion}
\label{sec:EoM_with_retardation}
In this section, we propose a heuristic modification of the elastic and self--inertial forces in which retarded effects are included. The retarded forces $F^\mathrm{II}_{\rm r}$ and $F^\mathrm{K}_{\rm r}$ are far much difficult to derive than  $F^\mathrm{II}_\mathrm{i}$ and $F^\mathrm{K}_\mathrm{i}$ because of the hypothesis of unsteady velocity and stress fields. For this reason no numerical estimation of these terms is given here. 

\subsection{Retarded inertial force}
To get rid of the assumption of the stationarity made in the first model, we use a self--inertial force that takes into account emission and propagation of waves accompanying changes of velocity of the dislocation \cite{Pillon_etal_07}. This solution is constructed by using the retarded self--inertial force ($F^{\rm SI}_{\rm r}$) produced at time $t$ by an instantaneous velocity jump from 0 to $v$ at time $\tau<t$ \cite{Clifton_Markenscoff_81}:  
\begin{equation}
\label{eq:fcm} F^{\rm SI}_{\rm r}(t-\tau,v)=\frac{g\bigl(v\bigr)}{t-\tau}\;\;.
\end{equation}
with $g$ a function that depends on dislocation character. The work done by this trailing force balances the increase of total energy due to the progressive updating of strain and of velocity fields from the solution at $v=0$ to the one at $v>0$. The retarded inertial force for any function $v(t)$ is constructed by summing all the contributions of the trailing forces $\delta F = [\partial F(t-\tau, v(\tau) ) / \partial v(\tau) ] \delta v(\tau)$ due to elementary velocity jumps $\delta v$ at $t=\tau$ which we assume to be a reasonable approximation of the trailing force due to a jump at $\tau$ from $v$ to $v+\delta v$: 
 \begin{equation}
 \label{eq:edm_prb_gen} 
 F^{\rm SI}_{\rm r} =
 \int_{-\infty}^t 
 \frac{g'\bigl(v(\tau)\bigr)}{t-\tau}\dot{v}(\tau) {\rm d}\tau
 \;\;.
 \end{equation}
This expression of the self-inertial force is singular at $\tau=t$ due to the assumption of point dislocation done in the original work \cite{Clifton_Markenscoff_81}. A regularization of the time--kernel has been proposed by Pillon \emph{et al.} \cite{Pillon_etal_07} to account for a core-size for the dislocation and the retarded-inertial force now reads:
\begin{equation}
\label{eq:edm_prb_gen} 
 F^{\rm SI}_{\rm r} =
\int_{-\infty}^t
\frac{g'\bigl(v(\tau)\bigr)}{[(t-\tau)^2+t_0^2]^{1/2}}\dot{v}(\tau) {\rm d}\tau
\;\;,
\end{equation}
where $t_0=\zeta_0/c_{\rm s}$ and $\zeta_0$ the dislocation core width at rest. In \cite{Pillon_etal_07} it has been underlined that this expression leads to extremely small inertial force (compared to the one given in equation~(\ref{eq:edm_hirth})) when one focuses on short times scales, and has a complex dependence with respect to time, according to the fact that the zone experiencing an energy evolution is progressively expanding from the vicinity of the dislocation to the whole solid. This non--local in time force represents the interaction of a dislocation with its own past motion. For interacting dislocations, the construction of a retarded elastic interaction force is now proposed. 

\subsection{Retarded elastic force}
\begin{figure}
\begin{center}
\includegraphics[width=12cm]{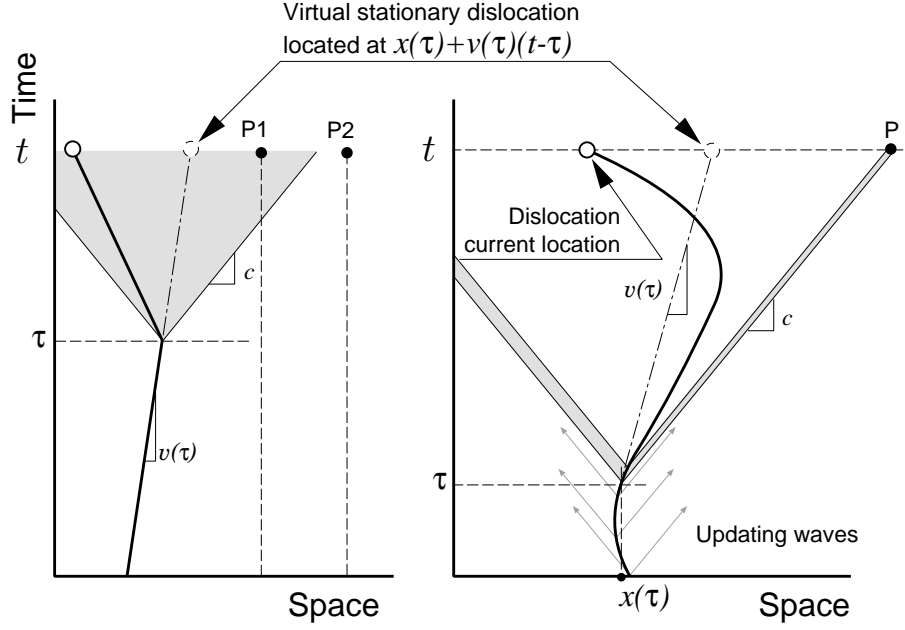}
\caption{\label{fig:champs_retard}Definition of the retarded elastic force. The solid curve represents the trajectory in a space-time diagram and the dashed-dots lines, the trajectory of the virtual dislocation which would have kept a stationary velocity. On the left-hand side is shown the delay induced by a jump from a stationary velocity $v$ to another velocity $v+\delta v$ and on the right-hand side, a general trajectory is illustrated.}
\end{center}
\end{figure}

The retarded elastic interaction force $F^\mathrm{E}_{\rm r}$ is due to stress around moving dislocations that progressively rearrange by wave propagation. We propose a construction of retarded elastic forces between moving dislocations. In the case of a single dislocation that jumps from $\vect{v}$ to $\vect{v}+\delta \vect{v}$ at instant $t=\tau$, the stress field at a distance $|\vect{p} - \vect{x}(\tau)|$ from the dislocation has been modified by the velocity jump $\delta \vect{v}$, provided that acoustic waves have been propagated up to this point, which is the case of any point $\vect{p}$ satisfying the condition $|\vect{p} - \vect{x}(\tau)| \leq c(t-\tau)$. We suppose that the stress at this point is equal to the stationary solution at velocity $\vect{v}+ \delta \vect{v}$ (this is the case of the point ``$P_1$" in the figure~\ref{fig:champs_retard}-left). Conversely, the field at distance $|\vect{p} - \vect{x}(\tau)| > c(t-\tau)$ from the source of the waves is not modified by the jump. It is therefore the field of a dislocation in stationary motion that \emph{ignores} the velocity jump, that is to say with a velocity $\vect{v}$ and located at $\vect{x}(\tau) + (t-\tau) \cdot \vect{v}$, different from the present position $\vect{x}(t)=\vect{x}(\tau) +  (t-\tau) \cdot (\vect{v}+\delta \vect{v})$. This second situation is the one of the point ``$P_2$" of the figure~\ref{fig:champs_retard}-left. Therefore, two stationary solutions have to be considered, depending on the relative location of the examination point (in ``$P_1$" or in ``$P_2$", see figure~\ref{fig:champs_retard}-left) where the stress is evaluated and of the dislocation trajectory. 

One can notice that the stress at $P_2$ is produced by a \emph{virtual} dislocation that, from the time $\tau$ has conserved the same stationary velocity $\vect{v}(\tau)$ (i.e. $\delta \vect{v}=0$) up to the present time $t$. In the case of a constant velocity the virtual dislocation is superposed to the real one and corresponds to the standard way to calculate interaction forces between moving dislocations \cite{Wang_etal_07,Roos_etal_01_a}.

For any dislocation motion, we suppose that a stationary solution is achieved at each time step but is visible only after the propagation of the waves and before any new change in the velocity, which will lead to a new stationary solution. Thus, the stationary solution to be considered at time $t$ and at the measurement point $\vect{p}$ is given by the element of trajectory in the past time $\tau$ for which the relation $|\vect{p} - \vect{x}(\tau)| = c(t-\tau)$ is verified, as depicted in figure~\ref{fig:champs_retard}-right by the gray zone.

In the case of two interacting dislocations of circulation $\vect{\xi}$, the Peach-Kohler interaction forces ($F^{E,\;\beta \rightarrow \alpha}_{\rm r}(t)$ coming from $\beta$ to $\alpha$ and conversely $F^{E,\;\alpha \rightarrow \beta}_{\rm r}(t)$ from $\alpha$ to $\beta$) is 
\begin{equation}
\label{eq:force_dislo_beta_sur_alpha_stat} 
\vect{F}^{E,\;\beta \rightarrow \alpha}_{\rm r}(t) =
\left\{
  \vect{\sigma}\left[\Delta\vect{p}^{\alpha\beta}, \; \vect{v}^{\beta}(\tau^{\beta}) \right]
  \cdot 
  \vect{b}^{\alpha}
\right\} \wedge \vect{\xi}^{\alpha}
\end{equation}
with $\wedge$ is the cross product, $\Delta\vect{p}^{\alpha\beta} = \vect{x}^{\alpha}(t) - [\vect{x}^{\beta}(\tau^{\beta})+ \vect{v}^{\beta}(\tau^{\beta}) \cdot(t-\tau^{\beta}) ]$ represents the relative location between the dislocation $\alpha$ and the virtual dislocation $\beta$ and $\vect{\sigma}\left[\Delta\vect{p}, \; \vect{v}\right]$ the stress tensor at $\Delta\vect{p}$ of a dislocation with a velocity $\vect{v}$. The past time $\tau^{\beta}$ at which the stationary solution is considered is given by the intersection of the trajectory $\vect{x}^{\beta}(\tau^{\beta})$ with the space--time cone $|\vect{x}^{\alpha}(t) - \vect{x}^{\beta}(\tau^{\beta})| = c(t-\tau^{\beta})$.

The force $F^{E, \;\alpha \rightarrow \beta}_{\rm r}(t)$ is obtained by a similar way but may result in time $\tau^{\alpha}$ and distance $\Delta\vect{p}^{\alpha\beta}$ that is distinct from $\tau^{\beta}$ and $\Delta\vect{p}^{\beta\alpha}$. This is different to the standard definition of interacting forces for which no propagation time are considered (that is $\tau^{\beta} = \tau^{\alpha} = t$), leading to an equality of the distances between dislocations (i.e. $\Delta\vect{p}^{\alpha\beta} = \Delta\vect{p}^{\beta\alpha}$).

In this simplified construction we use stationary solutions, which amount to the neglect of transient waves accompanying any changes in velocity. Removing this hypothesis is made possible by considering exact solutions, as for example in the method based on Green functions proposed by Mura \cite{Mura_63} for expanding loops. These exact solutions involve however an additional temporal integration of a prohibitive computational cost for DD simulations.
\section{Applications}
The EoMs of the previous section have been implemented in a two dimensional (point like) DD code for validation. Such validation is made by comparing the results of the DD simulation and those of a more fundamental nature called \emph{Peierls-Nabarro Galerkin} (PNG) simulation \cite{Denoual_04,Denoual_07}. Indeed, the latter technique allows for a full-dynamic description of dislocations interactions and has the advantage to share the same set of hypothesis than the ones used for the definition of the EoM, namely, an isotropic elasticity, a continuous representation of the solid and a simplified microscopic viscosity.

A first simulation (a dipolar interaction), dedicated to the analysis of retardation effects is proposed in which the coupling terms $F^\mathrm{K}_\mathrm{i}$ and $F^\mathrm{II}_\mathrm{i}$ are expected to play a minor role. A second simulation (a coplanar annihilation), which can be seen as a two-dimensional substitute for a more general study on junctions magnifies these coupling terms, with however a non-negligible influence of retardation mechanisms.

For each configuration, the two dislocations are supposed to be parallel edge dislocations of opposite signs. The line direction of the two dislocations are oriented along the $z$-direction and they have opposite Burgers vectors of magnitude $b$. They are separated along the $y$-direction by a distance $h$ and are restricted to glide in the $x$-direction. A viscosity is introduced on the glide plane for the PNG simulations and permits the existence a finite dislocations velocity for constant applied stress. This ``microscopic'' viscosity brings to a ``mesoscopic'' dislocation viscosity, following the model presented by Rosakis \cite{Rosakis_01} and is implemented in the DD technique. 
Since the aim of this paper is to analyse and to model retardation and inertial effects, whatever the model of viscosity considered, no attempt was made to introduce complex dissipation phenomena emerging from the interaction of the dislocation core with the atomic lattice \cite{Koizumi_etal_02,Marian_Caro_06,Lin_Gao_Gumbsch_08}. In the following subsection, we briefly describe the PNG methods and some new improvements developed to provide comparisons with DD simulations. 

\subsection{Reference PNG simulations}
The Peierls-Nabarro Galerkin (PNG) method \cite{Denoual_04,Denoual_07} is a generalization of the Peierls-Nabarro concept in which the displacement fields are represented by an element--free Galerkin method, close conceptually to the finite elements method. The non-linear behaviour is introduced by allowing a displacement jump $\eta$ along the glide plane at the cost of an additional energy $\gamma^\mathrm{isf}$ (the \emph{inelastic stacking fault} energy) that is deduced from the $\gamma-$surface \cite{Denoual_04}. Incorporation of kinetic energy allows for acoustic waves, which are essential for instationary dislocations motion (see, for example the modelling of accelerated dislocations \cite{Pillon_etal_07}). This method has shown to reproduce very well the analytical solutions for a stationary--moving dislocation, even in the high velocity (relativistic) regime and has been used by Pillon \emph{et al.} \cite{Pillon_etal_07} to check the EoM defined by equation~(\ref{eq:edm_prb_gen}) for a single dislocation.

All the DD simulations are done in an unbounded domain $\Omega$, naturally avoiding dislocations images \cite{Hirth_Lothe_82}. To prevent from dislocation images in the corresponding PNG simulation, the displacement fields on the boundary $\partial \Omega$ is given by the convolution product of displacement field of a point dislocation in stationary motion with the dislocation density $\nabla \eta(s)$
\begin{equation}
\label{eq:u_ref_cl_png}
\vect{u}^{\rm imp}(\vect{r}_{\partial \Omega}) =\int_{L} \nabla \eta(s) \vect{u}^{\rm stat}\left(\vect{r}_{\partial \Omega} - \vect{s}, v(s)\right) {\rm d}s 
\;\; ,
\end{equation}
where $L$ is the glide plane. The displacement field $\vect{u}^{\rm stat}\left(\vect{r}, v\right)$ is the exact relativistic displacement field \cite{Weertman_61} generated by stationary dislocation moving at the velocity $v(s)$. The straightforward choice for $v$ in equation~(\ref{eq:u_ref_cl_png}) should be the velocity of each infinitesimal dislocation $v =  \dot{\eta} / \nabla \eta$. Nevertheless, this definition induces a noisy measure of $v(s)$ which is transmitted to the boundary conditions and may generate acoustic waves. The interaction of these waves with the dislocation modifies the velocity $v(s)$ and can bring on increasing oscillations. To get rid of this possibly resonating behaviour, the velocity $v(s)$ is replaced by an average along the glide plane of the velocity of each infinitesimal dislocations $v = \left< \dot{\eta} / \nabla \eta \right>$. This averaged velocity $v$ is then filtered in time with a first-order filter $v^{\rm filt} + \tau_{\rm f} \dot{v}^{\rm filt} = v $ and is used to define the imposed displacement $\vect{u}^{\rm imp}$. The characteristic time $\tau_{\rm f}$ is set to the time needed to accelerate the dislocation up to stationary motion.

To be comparable with the DD simulations, PNG simulations of the interaction must be done with dislocations that are close to stationarity. To clean all the instationary information due to the initial acceleration of the dislocations, a body force proportional to the difference between a stationary velocity field $\dot{\vect{u}}^{\rm imp}$ and the present one $\dot{\vect{u}}$ is applied
\begin{equation}
\label{eq:png_amortissement} 
\mathrm{div} \left( \vect{\sigma} \right) - \rho \ddot{\vect{u}} 
=
\frac{\rho}{\tau_{\rm c}}\left[ \dot{\vect{u}}-\dot{\vect{u}}^{\rm imp}(\vect{r}_{\Omega})\right]
\;\; ,
\end{equation}
where $\tau_{\rm c}$ is a characteristic time and $\dot{\vect{u}}^{\rm imp}(\vect{r}_{\Omega})$ the field defined in equation~(\ref{eq:u_ref_cl_png}) in the whole volume $\Omega$. The convergence time towards the stationary solution is of the order of $\tau_{\rm c}$. 

This convergence mechanism is set to zero (ie $\tau_{\rm c}\rightarrow \infty$) at the beginning of the interaction so that the instationary solution is no longer altered. Thanks to this procedure, displacement and velocity fields are in accordance to stationary motion. The comparison with DD simulations presented in the following begins after this initialization step. 

 \subsection{Dipolar interaction}
The coupling terms $F^\mathrm{K}_\mathrm{i}$ and $F^\mathrm{II}_\mathrm{i}$ represent the difference of kinetic and inertial energy between interacting dislocations and the same dislocations considered as isolated ones. Preliminary simulations have shown these terms to decrease rapidly when the distance between dislocation increases. A dipolar interaction with a distance between gliding planes of several Burgers vectors is therefore a good configuration to test retardation effects, with a weak influence of coupling terms. Consequently, the kinetic interaction force and the inter-\-inertial force can be neglected in the dipolar interaction problem.

The initial trajectories and boundary conditions used in the DD and PNG simulations are the same with a steady state motion at $t=0$. The two dislocations move symmetrically and at each time, the retarded elastic force is calculated by searching in the past the element of trajectory defining the location and velocity of the virtual dislocation, as already described in section~\ref{sec:EoM_with_retardation}. The distance $h$ in the $y$-direction has been fixed in PNG and DD simulations to $8b$ which corresponds to an average value of minimum heights for dipole observed experimentally \cite{Veyssiere_07}. The cut off radius $R = 500\mu m$ (or $2500 b$) in equation~(\ref{eq:mass_hirth}) corresponds to a typical dislocation density of $10^{12}/m^2$. The influence of retarded elastic force, and of retarded inertial forces are analyzed by turning on and off $F^\mathrm{E}_\mathrm{i}$/$F^\mathrm{E}_{\rm r}$ and $F^\mathrm{SI}_\mathrm{i}$/$F^\mathrm{SI}_{\rm r}$. In $F^\mathrm{E}_{\rm r}$, the velocity $c$ is supposed to be the velocity of shear waves $c_{\rm S}$, in accordance with observations of Pillon \emph{et al.} \cite{Pillon_etal_07} where $c_{\rm S}$ brings the main  contribution to self--inertia. 

\begin{figure}
\begin{center}
\includegraphics[width=12cm]{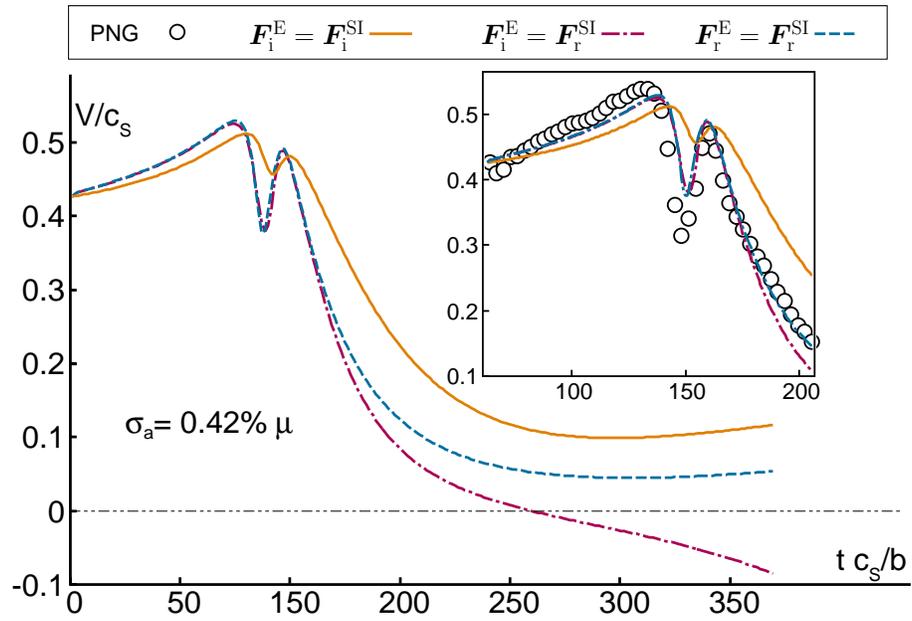}
\caption{\label{fig:courbes_comp_dipole} (Colour online). Velocities vs time for a dislocation interacting with another one in a dipole of edge dislocations. We first consider EoM without retardation ($F^\mathrm{E}_\mathrm{i} = F^\mathrm{SI}_{\rm i}$), then we add the retarded inertial force ($F^\mathrm{E}_\mathrm{i} = F^\mathrm{SI}_{\rm r}$) and finally we use fully retarded EoM ($F^\mathrm{E}_\mathrm{r} = F^\mathrm{SI}_{\rm r}$).
}
\end{center}
\end{figure}

Figure~\ref{fig:courbes_comp_dipole} shows results obtained by the PNG technique and the DD technique, the retarded aspect for the $F^\mathrm{E}$ and $F^\mathrm{SI}$ being turned on and off. The simulations are done for a constant applied stress $\sigma_{\rm a}$. During the short-range interaction, DD simulations show no influence of the retarded interaction force. This is consistent with a propagation time of about $8 b/c$ (or in dimensioned time $\approx 0.3$ ps) short compared to the characteristic time of the interaction ($\approx 50 b/c$, or in dimensioned time $\approx 1.7$ ps), inducing negligible retardation effect. This similarity progressively vanishes when the interaction distance increases, and the instant elastic force eventually gives a stable configuration ($v=0$) whereas the retarded force predicts a complete separation of the two dislocations. The corresponding PNG simulations cannot be achieved up to this time but clearly follows the retarded interaction simulation.

The difference between EoM with retarded self-inertial term and EoM with an instant mass is more contrasted during the short--range interaction, the better match with PNG simulations being obtained by the EoM with retarded self-inertial term. This discrepancy is due to an overestimation of the characteristic size of the zone playing a role in the instant mass. In the instant EoM, this size has been set to $R=2500 b$. However, most of the variations of the velocity take place in a time range of less than 50 $b /c_\mathrm{s}$, which limits the zone contributing to inertial effect to $50\times b$ (see figure~\ref{fig:courbes_comp_dipole}), leading to a strong overestimation of the inertial effect by the instant self-inertial force. Finally, the critical stress that breaks the dipole is found to be 0.66\%$\mu$ without inertial effect and reduced to 0.42\%$\mu$ by using the retarded EoM whereas the instant mass gives only 0.37\%$\mu$. 

Therefore, retarded inertia play a part mostly when two dislocations cross each other (or, in a more general case when a junction forms) and cannot be modelled accurately with an instant mass. The dipole formation occurs thus easier than expected with an instant mass EoM, which promotes the crossing over between dislocations by an overestimation of their kinetic energies. Conversely, the effect of retarded elastic force influences mainly the long range interactions, and intensifies as the distance between dislocations increases. This can potentially have a strong influence for high strain rate since the adjustment of the dislocations location is only perceived in their local environment. This is especially relevant for shock loadings in which the stress modification induced by dislocations motion should be confined behind the shock front.

\subsection{Annihilation}
The use of a coplanar annihilation in place of the dipolar interaction results in an important increase of the relative influence of the coupling forces $F^\mathrm{K}$ and $F^\mathrm{II}$. A stiffer variation of the elastic forces (notably when the dislocations are superposed), leads, furthermore, to a significant sensitivity to retardation effects. Surprisingly, this reaction, which can be considered as the strongest possible one \cite{Madec_etal_03}, is shown in the following to be also breakable by inertial effects. This configuration is therefore a more severe test of the proposed EoM than the dipolar one, with an additional difficulty coming from the impossibility to separate the influence of each of the forces, coupled and retarded.

An additional interaction force, specific to coplanar annihilation, comes from the possibility to \emph{superpose} the dislocation, by the way modifying the overall energy stored into the dislocation core. Indeed, in the framework of the Peierls-Nabarro Galerkin method, the energy of a single isolated dislocation contains a core energy $\int_x \gamma^{\text{isf}} \left[ \eta(x) \right] \mathrm{d}x$ with $\gamma^{\text{isf}}$ a surface potential and $\eta(x)$ the displacement jump along the glide plane. This energy is constant for dislocation with a fixed core width and its influence is usually neglected. However, this energy completely vanishes when two dislocations of opposite sign superimpose. The corresponding potential energy is obtained by summing the displacement jump of the two dislocations :
\begin{equation}
\gamma^{\alpha\beta} =\int_S \gamma^{\text{isf}}  \left[ \eta(x-x^\alpha) -\eta(x-x^\beta)  \right] \mathrm{d} S
\end{equation}
where $\gamma^{\text{isf}}$ is the interplanar potential used in the PNG method and where $x^\alpha$ and $x^\beta$ are the dislocation locations. A rough estimation of this interaction force consists in taking for $\eta(x)$ the quasi-static solution of the dislocation displacement jump. An additional force $F^{\gamma}=-\partial \gamma^{\alpha\beta}/\partial x^\alpha$ derives from this potential, and appears in the left-hand side of equation~(\ref{e:EOM_synthetic}). Therefore, both influence of coupling forces, retardation effects and core energy could have an influence in this configuration and will be measured. 
 
We extend the method used for dipoles to coplanar annihilation by fixing the distance between glide planes to $h = 0$ and by introducing a finite core size to avoid singularity in the elastic force when the two dislocations meet. Contrary to the dipolar interaction, we measure the minimal applied stress above which dislocations renucleate after annihilation. The overall result depends potentially on coupling terms, retardation effects and core energy. The influence of the coupling forces $F^\mathrm{K}_\mathrm{i}$ and $F^\mathrm{II}_\mathrm{i}$ is tested by using the ``instant'' EoM and by turning them on and off. The role of retardation effects is estimated by comparing the retarded EoM (in which no coupling forces are known) and the instant one, in which $F^\mathrm{K}_\mathrm{i}$ and $F^\mathrm{II}_\mathrm{i}$ are switched off. The influence of core energy $F^{\gamma}$ will be studied in the two equations of motion.

\begin{figure}[tbh]
\begin{center}
\includegraphics[width=10cm]{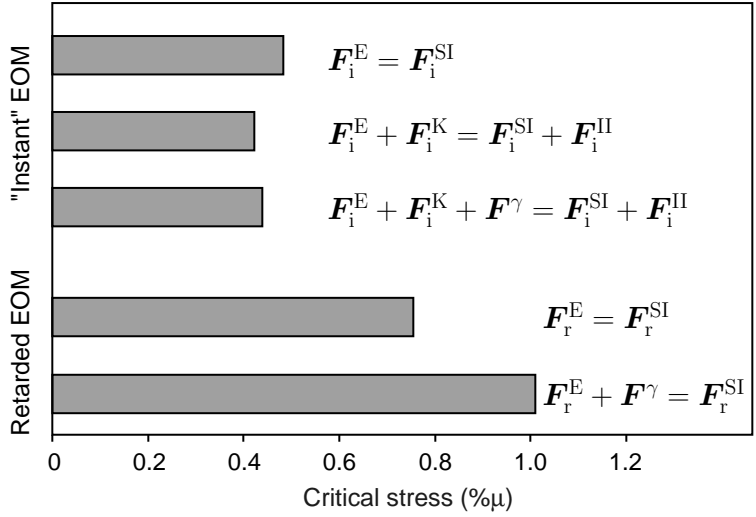}
\caption{\label{fig:resultats_relatifs_annihilation} Summary of results obtained for the coplanar annihilation.}
\end{center}
\end{figure}

Results obtained in each cases are summarized in figure~\ref{fig:resultats_relatifs_annihilation}. The $F^\gamma$ is found to modify only the retarded EoM. This is consistent with an evolution of this force only when the dislocation cores are in contact. This force is therefore a very brief signal that is felt \emph{only} with the retarded EoM which is known to predict low inertia for high frequency loadings \cite{Pillon_etal_07}. 

Contrary to the force $F^\gamma$, the introduction of $F^{\mathrm{K}}$ and $F^{\mathrm{II}}$ in the ``instant'' EoM decreases slightly the critical applied stress. Indeed, the energy $k^{\alpha\beta}$ is positive and has its maximum when $e^{\alpha\beta}$ is minimum. The kinetic interaction force partially compensates the elastic interaction forces, which in turn decreases the critical stress. The force $F^{\mathrm{II}}$ has approximately the same influence as $F^{\mathrm{K}}$. Actually, for low velocity dislocation the derivative of $e^{\alpha\alpha} + e^{\alpha \beta}$ with respect to the velocity is very low (the elastic field weakly depends on velocity) and only the energy $k^{\alpha\alpha} + k^{\alpha\beta}$ intervenes in the mass. For the same reason as before, $k^{\alpha\alpha} + k^{\alpha\beta}$ is intensifying during annihilation, by the way increasing the inertial energy which helps the crossing of the dislocations. 

The critical stresses given by the proposed EoMs are however far from the results of PNG simulations that predict a crossing for applied stresses of $2.7\%\mu$ but a complete annihilation for $2.4\%\mu$. An explanation of this discrepancy is now proposed. 

\section{Discussion}
\label{discussion}

Quasi-static reactions between dislocations are known to be correctly described as lines in elastic interaction without any core contribution \cite{Rodney_Phillips_99, Madec_etal_03}. It is striking to notice that, in opposition to the quasi-static case, dynamic annihilation cannot be quantitatively modelled by the proposed EoM, even by taking care of wave propagation through retarded mechanisms. The dislocations energy is therefore damped down by another mechanism that cannot be represented in terms of superimposition of two dislocations.

\begin{figure}[t]
\includegraphics[width=6cm]{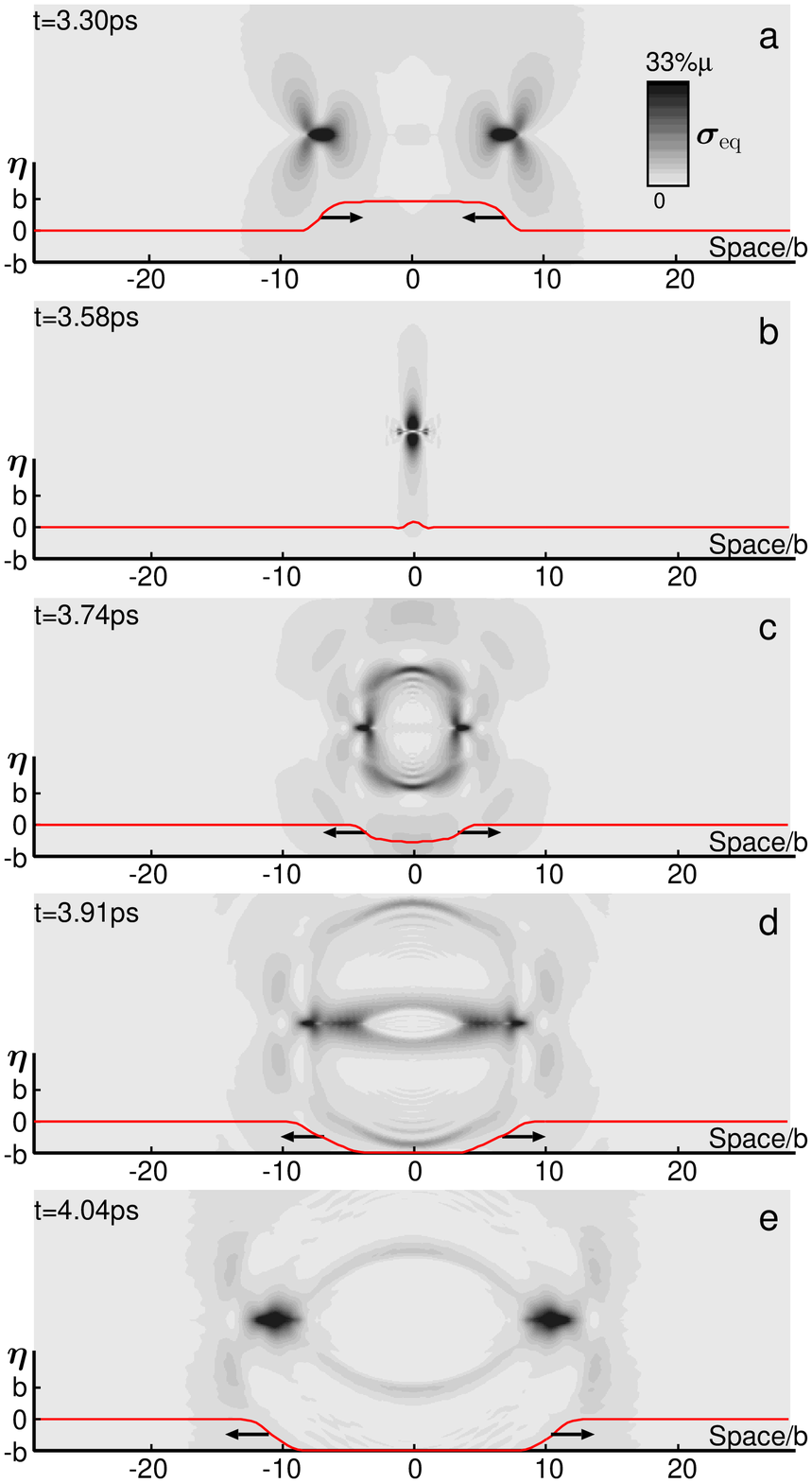} \includegraphics[width=6cm]{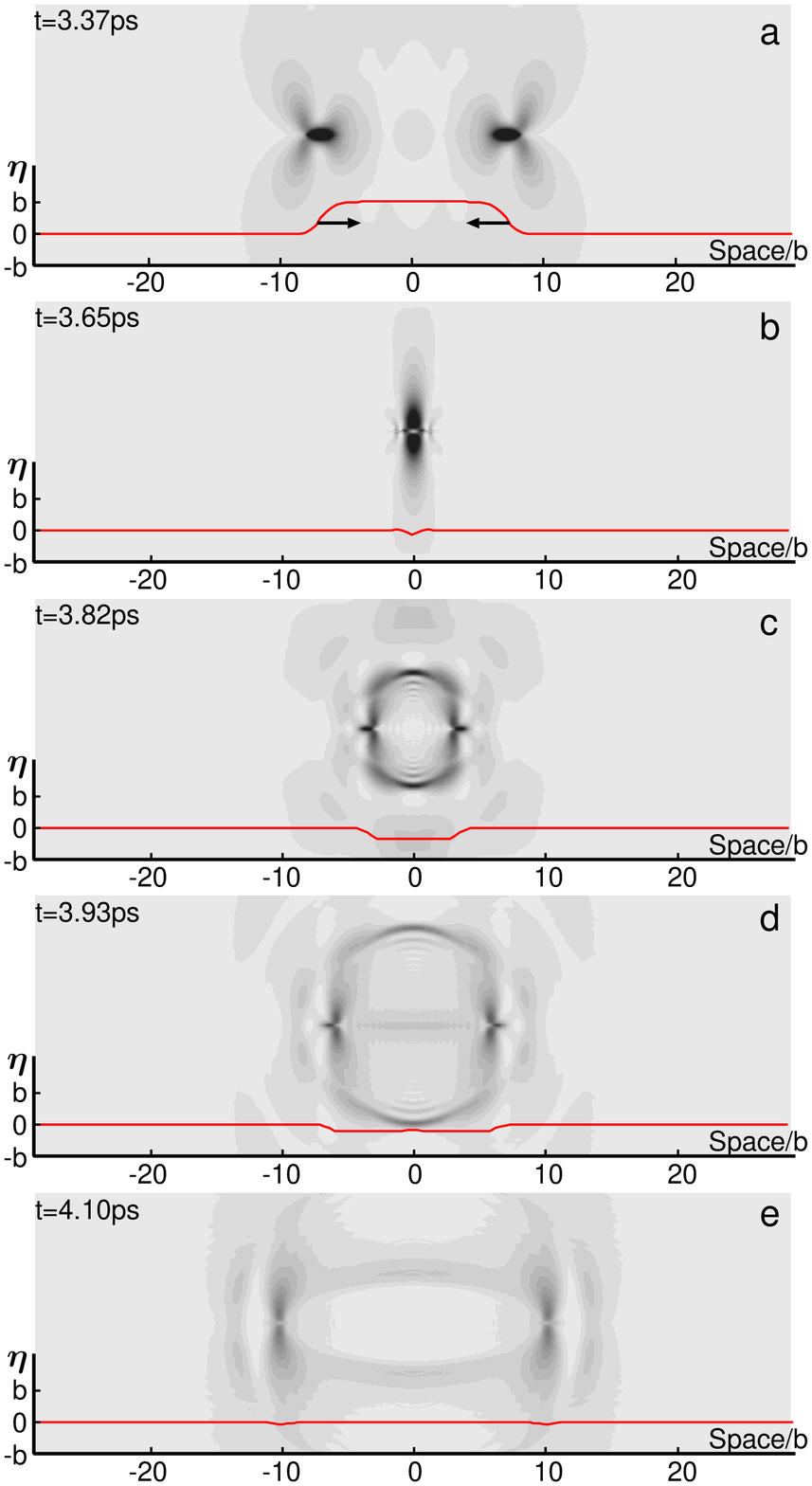}

\caption{\label{fig:PNG}(Color online). (Left) Snapshots of PNG simulations of a coplanar interaction with $2.7\%\mu$ as applied stress. Pictures represent the equivalent stress in the whole space with in addition the corresponding displacement jump on the glide plane. (Right) The same simulation with $2.4\%\mu$ as applied stress. }
\end{figure}

\begin{figure}[t]
\begin{center}
\includegraphics[width=8cm]{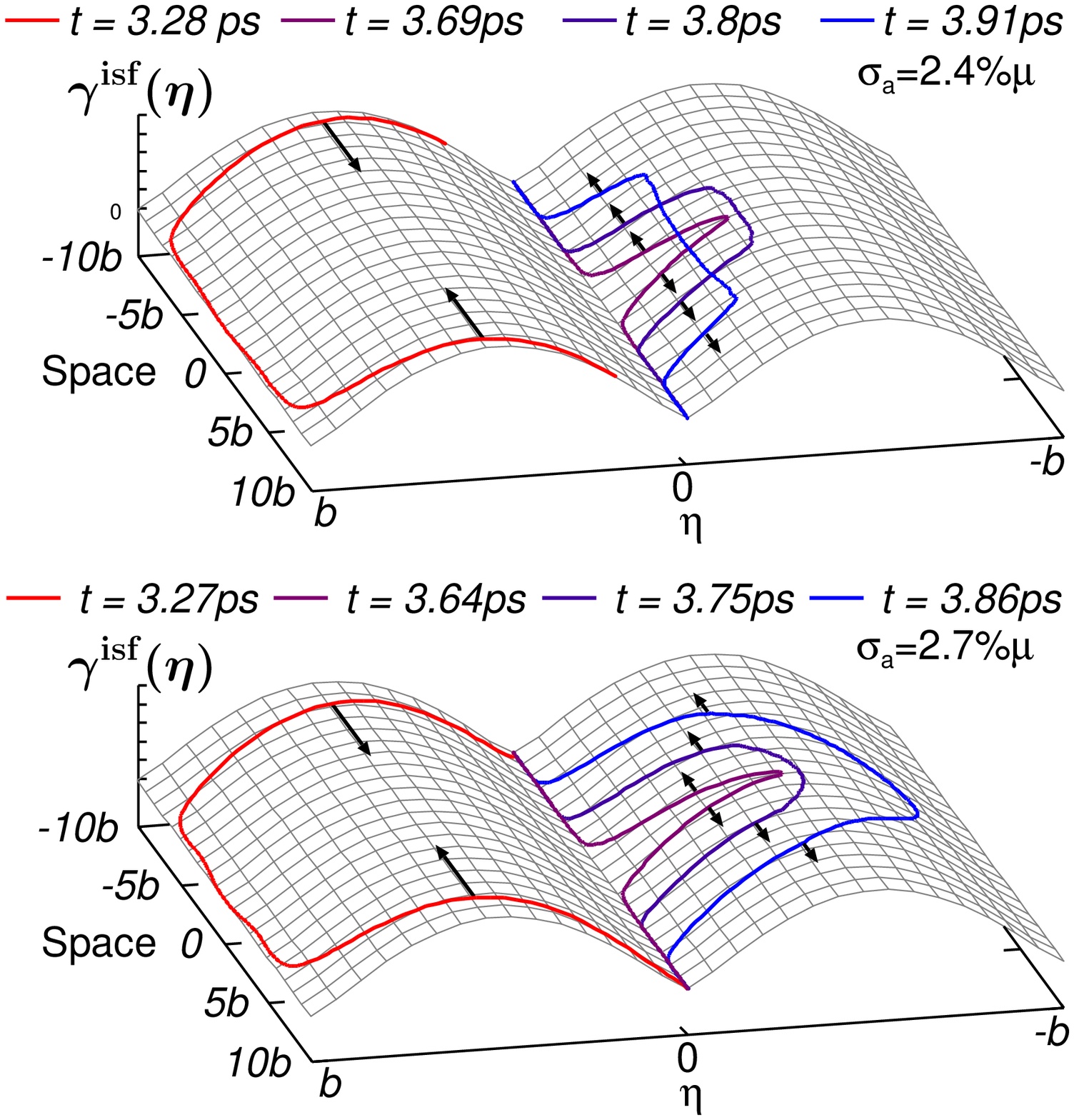}
\caption{\label{fig:potentiel_eta}(Color online). Evolution of the inelastic stacking potential $\gamma^{\text{isf}}(\eta)$ along the displacement jump $\eta$ and along the slip plane for different times. The two configurations corresponding to an applied stress of $\sigma = 2.4\%\mu$ and $\sigma = 2.7\%\mu$ have been represented.}
\end{center}
\end{figure}

The difference between analytical modelling and PNG simulations can be explained by analyzing the role played by the interplanar potential $\gamma^{\rm isf}$ during the reaction (figures~\ref{fig:PNG} and \ref{fig:potentiel_eta}). In figure~\ref{fig:PNG}--a, the two dislocations are getting closer at a velocity of $\approx 0.83 c_{\rm s}$ and eventually superpose (figure~\ref{fig:PNG}--b). At this time, no dislocation is present (i.e. $\eta \approx 0$ everywhere) and the energy is mainly of kinetic nature. The available amount of energy results in the creation of two dislocations of magnitude half the initial Burgers vector in figure~\ref{fig:PNG}--c. This two ``partial'' dislocations, separated by more than 4 $b$, are thus accumulating, as a staking fault, an important amount of potential energy. This state is very different from the expected ``ideal'' entire dislocations (half of each dislocations is missing) and changes dramatically the displacement fields that induces an intense stress wave (figure~\ref{fig:PNG}--c). This state is however unstable since the $\gamma^{\rm isf}$ does not have any local minimum at $b/2$ (see figure~\ref{fig:potentiel_eta}): the field $\eta$ has to go through this potential barrier to achieve $\eta=b$ or to turn back to $\eta=0$.

For an applied stress of $2.4\%\mu$, kinetic energy is not high enough to allow the dislocation to overcome the maximum of the inelastic stacking fault energy at $\eta = b/2$. This is illustrated on figure~\ref{fig:potentiel_eta}, where the inelastic staking fault is plotted with respect to the position on the slip plane and to the value of the displacement jump $\eta$. Since the available kinetic energy is not sufficient to overcome the barrier, the displacement field $\eta$ minimizes the potential energy by decreasing to a homogeneous value of $\eta=0$ along the slip plane. To compensate this decrease of the inelastic stacking fault, a second wave is emitted in the solid (see figure~\ref{fig:PNG}--right-e). Conversely, for an applied stress of $2.7\%\mu$, kinetic energy is large enough to overcome the potential maximum for $\eta = b/2$ and to create two dislocations (with a displacement jump ranging from $\eta = 0$ to $\eta =-b$). The sudden change from two half dislocations to two entire ones induces the emission of a wave in the solid (figure~\ref{fig:PNG}--left--d and --e). The ``renucleated" dislocations moves with a velocity of $\approx 0.78 c_{\rm s}$.

Therefore, most of the available energy is dissipated in a ``two steps mechanism''. A first step consists in storing an important part of the kinetic energy into staking fault energy while another part is lost by an intense acoustic emission.  The second step results in a complete relief of the staking fault energy, leading one more time to an important acoustic wave emission. This scheme, very different from a simple superposition of dislocations, explains the discrepancy between PNG and the EoMs. Such phenomenon could also \emph{a priori} intervene during other contact reactions like junctions formation, as far as junction breaking due to inertia can be invoked.

\section{Concluding remarks}
An equation of motion for interacting dislocations is proposed by using two descriptions for stress and velocity fields, denoted ``instant" and ``retarded". A comprehensive study of the forces acting during dislocations interaction shows that in addition to the usual elastic and inertial terms, a kinetic interaction force and an inter-inertial force should be considered in EoMs for fast interacting dislocations. 

We show that inter--inertial force and inter--kinetic force does not play a significant role during short distance interaction like formation of dipoles, as far as an ``instant'' EoM can be considered. The retarded effects introduced in the inertial terms become important for contact reactions like annihilation and possibly formation of junctions. For instance, the ``retarded'' EoM leads to critical stresses for dislocation annihilation two times higher than the ``instant'' EoM. 

In addition, we show that the elastic retarded force modifies long range interactions and therefore is essential to model shock loadings. In such conditions, moving dislocations will concentrate stress \emph{only} behind the shock front and as a consequence nucleate original plastic features.
 
The comparisons between DD and PNG simulations shows however that ``instant'' as well as ``retarded'' EoMs, are both failing to reproduce quantitatively the inertial effects observed at the limit case of coplanar annihilation. In such case, an original mechanism of energy accumulation into the interplanar potential is proven to be the reason for DD simulations deficiency. More generally, we show that inertial effects can strongly influence contact reactions. As an example, two dislocations with opposite Burgers vector can completely annihilate and renucleate as a result of inertia. From this observation, related to the most energetically favourable dislocation-dislocation reaction, one can conclude that inertia may be determinant in many strain-hardening mechanisms involved during high-strain rate loadings.

\section*{Acknowledgments}
The authors gratefully thank Y.P. Pellegrini for insightful discussions about equations of motion, B.~Devincre for precious comments on this manuscript and R. Madec for discussions.



\end{document}